\documentclass[11pt]{article}   
\usepackage{graphicx,floatflt,amssymb}    
\usepackage{epsfig,axodraw}     
\usepackage{graphics}     
\usepackage{psfrag}     
\newcommand{\ba}{\begin{array}}      
\newcommand{\ea}{\end{array}}      
\newcommand{\bd}{\begin{displaymath}}      
\newcommand{\ed}{\end{displaymath}}      
\newcommand{\be}{\begin{equation}}      
\newcommand{\ee}{\end{equation}}      
\newcommand{\bea}{\begin{eqnarray}}      
\newcommand{\eea}{\end{eqnarray}}

\newcommand{\rpv}{\mbox{$\not \hspace{-0.10cm} R_p$ }}

\def\Journal#1#2#3#4{{#1} {\bf #2}, #3 (#4)}

\def\NPB{{\em Nucl. Phys.} B}  
\def\PLB{{\em Phys. Lett.}  B}  
\def\PRL{\em Phys. Rev. Lett.}  
\def\PRD{{\em Phys. Rev.} D}  
\def\ZPC{{\em Z. Phys.} C}  
\def\PR{\em Phys. Rep.}   
 
\def\JHEP{\em JHEP}

\begin{document}      
\vspace*{-0.5in}      
\renewcommand{\thefootnote}{\fnsymbol{footnote}}      
\begin{flushright} 
LPT Orsay/06-39 \\ 
SINP/TNP/06-14\\            
\texttt{hep-ph/0606179}       
\end{flushright}      
\vskip 5pt

\begin{center}      
{\Large {\bf 
A new mechanism of neutrino mass generation in the NMSSM \\
with broken lepton number 
}} \\
\vspace*{1cm}
{\sf Asmaa Abada $^{1,}$\footnote{E-mail 
 address: {\tt abada@th.u-psud.fr}}},  
{\sf Gautam Bhattacharyya  $^{2,}$\footnote{E-mail 
 address: {\tt gb@theory.saha.ernet.in}}}, 
{\sf Gr\'egory Moreau $^{3,}$\footnote{E-mail 
 address: {\tt greg@cftp.ist.utl.pt}}} 

\vspace{10pt} 
{\small $^1${\it Laboratoire de Physique Th\'eorique, 
Universit\'e de Paris-sud XI \\   
B\^atiment 210, 91405 Orsay Cedex, France} \\
$^2${\it Saha Institute of Nuclear Physics,    
1/AF Bidhan Nagar, Kolkata 700064, India} \\
$^3${\it Centro de F\'{\i}sica Te\'orica de Part\'{\i}culas, 
Departamento de F\'{\i}sica \\
Instituto Superior T\'{e}cnico, 
Av. Rovisco Pais 1, 1049-001 Lisboa, Portugal} 
}  
 
\normalsize      
\end{center} 
\vskip 10pt
 
\begin{abstract}   
In the Minimal Supersymmetric Standard Model with bilinear $R$-parity
violation, only one neutrino eigenstate acquires a mass at tree level,
consequently experimental data on neutrinos cannot be accommodated at
tree level.  We show that in the Next-to-Minimal extension, where a
gauge singlet superfield is added to primarily address the so-called
$\mu$-problem, it is possible to generate two massive neutrino states
at tree level.  Hence, the global three-flavour neutrino data can be
reproduced at tree level, without appealing to loop dynamics which is
vulnerable to model-dependent uncertainties. We give analytical
expressions for the neutrino mass eigenvalues and present examples of
realistic parameter choices.
\vskip 0.4cm \noindent   
\texttt{PACS Ns:~ 11.30.Fs, 12.60.Jv, 14.60.Pq, 14.80.Ly} \\   
\texttt{Keywords:~ Neutrino mass, $R$-parity violation, NMSSM}   
\end{abstract}

\renewcommand{\thesection}{\Roman{section}}
\setcounter{footnote}{0}
\renewcommand{\thefootnote}{\arabic{footnote}}

\section{Introduction} 
\label{Introduction}
In the Minimal Supersymmetric Standard Model (MSSM) with bilinear
interactions in the superpotential explicitly violating the $R$-parity
symmetry \cite{rpar1}, a neutrino Majorana mass can be
generated. Nevertheless, the rank 1 nature of the neutrino mass matrix
suggests that only one eigenstate becomes massive at tree level,
whereas the neutrino oscillation data require at least two non-zero
mass eigenvalues \cite{Valle}.  The bilinear $R$-parity violating
(\rpv) soft masses induce one more non-zero eigenvalue, but only at
one-loop order. What happens if one considers the next-to-minimal
version of the MSSM, called `NMSSM' \cite{NMSSM}, in the presence of
bilinear \rpv terms in the superpotential ?  Here, the particle
content is extended by one gauge singlet superfield. Our main result
in this work is that {\em two} non-degenerate massive neutrino states
now emerge at {\em tree level}. The upshot is that one can now
reproduce the neutrino oscillation data with the superpotential
parameters and gaugino masses just from {\em tree level} physics. On
the other hand, in the usual MSSM with bilinear \rpv terms, the
generation of the second neutrino mass eigenvalue relies on the soft
supersymmetry breaking scalar masses which in turn bring more
uncertainties from the supersymmetry breaking mechanism; furthermore,
uncertainties from loop dynamics creep in too.

An increasingly important virtue of the NMSSM \cite{NMSSM} (see
\cite{NMpheno} for phenomenological studies) is that it ameliorates
the `little hierarchy' problem originating from the requirement of
large soft supersymmetry breaking scalar masses compared to the
electroweak scale (in order to sufficiently push the lightest Higgs
mass beyond the LEP limit). The NMSSM also provides a solution to the
so-called $\mu$-problem by arranging the vacuum expectation value
(vev) of the gauge singlet scalar of the order of the supersymmetry
breaking scale, so that the $\mu$ parameter turns out to be at the
electroweak scale.

A recent paper \cite{rpNMSSM1}, in the context of NMSSM with \rpv
couplings, deals with the generation of neutrino masses where two
eigenvalues arise at loop level. In another recent analysis
\cite{rpNMSSM2}, it has been shown that the NMSSM with bilinear \rpv
terms offers a possible mechanism of neutrino mass suppression, thus
significantly reducing the hierarchy between $\mu_i$ and $\mu$
(defined below).  Besides, an alternative extension of the MSSM
explicitly breaking the $R$-parity has been proposed in order to
simultaneously address the $\mu$-problem and the issue of small
neutrino masses \cite{Munoz}.

In Section \ref{Superpotential}, we present the superpotential of the
model we consider.  In Section \ref{Matrix}, we discuss the effective
neutrino mass matrix. We present our numerical results in Section
\ref{Numeric}. Finally, we conclude in Section \ref{Conclusion}.

\section{Superpotential}
\label{Superpotential}
The NMSSM superpotential contains two dimensionless couplings $\lambda$ and
$\kappa$ in addition to the usual Yukawa couplings:
\begin{equation} 
W_{\rm NMSSM}= 
Y^u_{ij} Q_i H_u U_j^c + Y^d_{ij} Q_i H_d D_j^c + Y^\ell_{ij} L_i
H_d E_j^c + \lambda S H_u H_d + \frac{1}{3} \kappa S^3 \,
\label{WNMSSM} 
\end{equation} 
where $Y^{u,d,\ell}_{ij}$ are the Yukawa coupling constants ($i,j,k$ are
family indexes), and $Q_i$, $L_i$, $U^c_i$, $D^c_i$, $E^c_i$, $H_u$, $H_d$,
$S$ respectively are the superfields for the quark doublets, lepton doublets,
up-type anti-quarks, down-type anti-quarks, anti-leptons, up Higgs, down
Higgs, extra singlet under the standard model gauge group.
An effective $\mu$ term, given by $\lambda \langle s \rangle H_u H_d$, 
is generated via the vev of the scalar component $s$ of the singlet
superfield $S$.

We now take note that in supersymmetric theories there is no deep
underlying theoretical principle for the imposition of $R$-parity as a
symmetry
\cite{PhysRep}.  However, there exist strong constraints on the \rpv couplings
coming from various phenomenological considerations \cite{REVbounds,bounds}.
Limits on neutrino masses and mixings have also been translated into tight
upper bounds for \rpv couplings \cite{nubounds}.

In the present paper, we consider a generic NMSSM superpotential
containing both the bilinear and trilinear \rpv terms:
\begin{equation} 
W = W_{\rm NMSSM} + \mu_i L_i H_u + \lambda_i S L_i H_u , 
\label{Wgeneric} 
\end{equation} 
where $\mu_i$ ($\lambda_i$) are the dimension-one (dimensionless) \rpv
parameters.  Actually, the contribution of trilinear term $\lambda_i S
L_i H_u$ was studied in Ref.~\cite{rpNMSSM1}.  Admittedly, the most
generic NMSSM superpotential also contains the other renormalizable
trilinear \rpv interactions, namely, $\lambda_{ijk} L_i L_j E_k^c$,
$\lambda'_{ijk} L_i Q_j D_k^c$ and $\lambda''_{ijk} U^c_i D^c_j
D_k^c$, which are not relevant so long as we stick to tree level
neutrino mass matrix.

Normally, in the NMSSM, only trilinear couplings with dimensionless parameters
(like $\lambda$ and $\kappa$) are kept in the superpotential, while
dimensional parameters (like $\mu$) are generated from the vev $\langle s
\rangle$. In this paper, the \rpv NMSSM superpotential (\ref{Wgeneric}) is
assumed to arise in either one of the following three possible scenarios:
\begin{enumerate} 
\item 
All possible renormalizable terms are included in the
superpotential. Then both bilinear ($\mu_0 H_u H_d$, $\mu_i L_i H_u$)
and trilinear ($\lambda S H_d H_u$, $\lambda_i S L_i H_u$) terms are
admitted.  However, even if one may start with a term $\mu_0 H_u H_d$,
it can be rotated away by a redefinition of fields through a rotation
on $L_\alpha=(H_d,L_i)$ [$\alpha=0,\dots,3$], since $H_d$ and $L_i$
have the same gauge quantum numbers.  There is no reason why this
redefinition would remove also the $\lambda_\alpha$ terms
($\lambda_\alpha=(\lambda,\lambda_i)$), since the corresponding $4
\times 4$ rotation matrix depends on the $\mu_\alpha$ parameters (the
generic case is considered here, where $\mu_\alpha$ and
$\lambda_\alpha$ are {\em not} proportional).  The coefficient $\mu_s$
of the $S^2$ term is assumed to be zero which can be considered as a
possible natural value for a superpotential parameter.  It should be
noted that in the standard NMSSM with conserved $R$-parity there is an
accidental $Z_3$ discrete symmetry whose spontaneous breaking causes
the domain wall problem.  In our version of the NMSSM, the
\rpv bilinear $L_i H_u$ term explicitly breaks that $Z_3$ symmetry.  
In this scenario, we simply assume the existence of the dimensionful
$\mu_i$ terms, which as we will see later will be constrained from
neutrino data. But we do not advance any theoretical reason as to why
$\mu_i$ would be small.

\item 
Our second scenario is based on the 't Hooft criteria of naturality:
the parameters $\mu_i$, $\mu_0$ and $\mu_s$ are naturally small if the
symmetry of the theory increases as these parameters are set to zero.
For instance, one can assume that somehow a weak breaking (compared to
the electroweak scale $Q_{\rm EW}$) of some symmetry (like e.g. a U(1)
symmetry forbidding the bilinear terms) generates the bilinear terms
in the superpotential associated to $\mu_i$, $\mu_0$, $\mu_s \ll
Q_{\rm EW}$. This small breaking would allow to address the
$\mu$-problem.  Indeed, the main contribution to the dimension-one
coefficient of $H_u H_d$ here comes from $\lambda \langle s
\rangle$, as $\mu_0 \ll \mu = \lambda \langle s \rangle \sim Q_{\rm
EW}$. The weak breaking of the symmetry is also responsible for the
smallness of \rpv couplings and neutrino masses, since $\mu_i \ll
\mu$. Thus, in such a scenario, the treatment of the $mu$-naturalness
({\em \`a la} NMSSM) and of the neutrino masses ({\em \`a la} \rpv)
are nicely connected via the weak breaking of a common symmetry.
Admittedly, we do not provide any specific realisation of this weak
breaking.  We only hint at such a possibility that the bilinear
$\mu_i, \mu_0, \mu_s$ couplings may arise from powers of some small
spurion vev ($\ll Q_{\rm EW}$).

\item 
Finally, we propose a scenario where the trilinear $\lambda_i$ terms
in superpotential (\ref{Wgeneric}) are not present. This scenario
relies on the $Z_3$ symmetry, where the chiral superfields transform
by exp($i 2\pi q/3$), with the following charge assignments: $q = 0$
for $U^c$, $D^c$, $E^c$; $q = 1$ for $S$, $H_u$, $H_d$; and $q = 2$
for $Q$, $L$. Such a symmetry allows all couplings in
Eqs.~(\ref{WNMSSM}) and (\ref{Wgeneric}) except $S L_i H_u$. The other
terms $S^2$ and $H_uH_d$ are also eliminated by this symmetry. A
spontaneous breaking of this $Z_3$ symmetry admittedly creates the
domain wall problem, as happens in the standard NMSSM {\em with}
$R$-parity. In this scenario, the $\mu$ term is created from the vev
of $S$, but the $\mu_i$ terms are present in the superpotential from
the beginning only to be constrained by neutrino data.
\end{enumerate}

\section{Neutrino mass matrix}  
\label{Matrix}
{\bf Neutralino mass matrix:} 
Within our framework, the neutralino mass terms read as, 
\bea 
{\cal L}^m_{\tilde \chi^0}=
-\frac{1}{2} \Psi^{0^T} {\cal M}_{\tilde \chi^0} \Psi^0 + {\rm h.c.}  
\label{LAGmass} 
\eea 
in the basis $\Psi^{0^T}$ $\equiv$ $(\tilde B^0, \tilde W^0_3, \tilde
h^0_d, \tilde h^0_u, \tilde s, \nu_i)^T$, where $\tilde h^0_{u,d}$
($\tilde s$) are the fermionic components of the superfields
$H_{u,d}^0$ ($S$) and $\nu_i$ [$i=1,2,3$] denote the neutrinos.  In
Eq.~(\ref{LAGmass}), the neutralino mass matrix is given, in a generic
basis (where $\langle \tilde \nu_i \rangle \equiv v_i \neq 0$, $\mu_i
\neq 0$ and $\lambda_i \neq 0$), by
\bea 
{\cal M}_{\tilde \chi^0} =  
\left(  
\begin{array}{cc} 
{\cal M}_{\rm NMSSM}  & \xi_{\rpv}^{T} \\ 
\xi_{\rpv} & {\bf 0}_{3 \times 3} 
\end{array} 
\right) ,
\label{CHImass} 
\eea 
where ${\cal M}_{\rm NMSSM}$ is the neutralino mass matrix
corresponding to the NMSSM. While writing the latter mass matrix, we
assume $v_i \ll v_{u,d}$ (the exact expression of $M_Z$ being given by
$v^2 = v_u^2+v_d^2+ \sum^3_{i=1} v_i^2 = 2c_{\theta_W}^2M_Z^2/g^2
\simeq (246/\sqrt{2} ~\mbox{GeV})^2$). Also, we use $s$ and $c$ to
stand for sine and cosine, respectively.
\bea 
{\cal M}_{\rm NMSSM} 
= 
\left( 
\begin{array}{cccccc} 
M_{1}&0 & -M_Z \ s_{\theta_W} \ c_\beta & M_Z \ s_{\theta_W} \ s_\beta & 0\\  
0 & M_{2} & M_Z \ c_{\theta_W} \ c_\beta & -M_Z \ c_{\theta_W} \ s_\beta  &0\\ 
-M_Z \ s_{\theta_W} \ c_\beta & M_Z \ c_{\theta_W} \ c_\beta & 0 &-\mu& 
-\lambda v_u\\ 
M_Z \ s_{\theta_W} \ s_\beta &-M_Z \ c_{\theta_W} \ s_\beta &-\mu&0& 
-\lambda v_d+\sum^3_{i=1} \lambda_i v_i \\ 
0&0&-\lambda v_u &-\lambda v_d+\sum^3_{i=1} \lambda_i v_i & 2 
\kappa \langle s \rangle  \\ 
\end{array} 
\right) . 
\label{NMSSMmass} 
\eea 
Above, $M_1$ ($M_2$) is the soft supersymmetry breaking mass of the bino 
(wino), $\tan \beta=v_u/v_d=\langle h^0_u \rangle /\langle h^0_d \rangle$, 
and $\mu = \lambda \langle s \rangle$. 
We assume for simplicity that $\lambda$, $\kappa$ and the soft supersymmetry  
breaking parameters are all real. 

In Eq.~(\ref{CHImass}), $\xi_{\rpv}$ is the \rpv part of the matrix mixing 
neutrinos and neutralinos: 
\bea 
\xi_{\rpv} =  
\left(  
\begin{array}{ccccc} 
-\frac{g'v_1}{\sqrt{2}} & \frac{gv_1}{\sqrt{2}} & 0 & \mu_1 + 
\lambda_1 \langle s \rangle & \lambda_1 v_u  \\ 
-\frac{g'v_2}{\sqrt{2}} & \frac{gv_2}{\sqrt{2}} & 0 & \mu_2 + 
\lambda_2 \langle s \rangle & \lambda_2 v_u  \\ 
-\frac{g'v_3}{\sqrt{2}} & \frac{gv_3}{\sqrt{2}} & 0 & \mu_3 + 
\lambda_3 \langle s \rangle & \lambda_3 v_u   
\end{array} 
\right). 
\label{RPVmass} 
\eea 
$g$ and $g'$ being the SU(2) and U(1) gauge couplings.

{\bf Effective neutrino mass matrix:} 
We restrict ourselves to the
situation where $v_i/v_{u,d} \ll 1$ (as before), $|\mu_i/\mu| \ll 1$ and
$|\lambda_i/\lambda| \ll 1$ so that (i) no considerable modifications
of the NMSSM scalar potential are induced by the additional bilinear
and trilinear term in superpotential (\ref{Wgeneric}), (ii) the
neutrino-neutralino mixing is suppressed, leading to sufficiently
small neutrino masses as shown later, and (iii) the effective neutrino
mass matrix can be written to a good approximation by the following
see-saw type structure,
\begin{equation} 
m_{\nu} = - \xi_{\rpv} \ {\cal M}_{\rm NMSSM}^{-1} \ \xi_{\rpv}^T. 
\label{seesaw} 
\end{equation} 
From Eqs.~(\ref{NMSSMmass}), (\ref{RPVmass}) and (\ref{seesaw}), we deduce
the analytical expression of the effective Majorana neutrino mass matrix:
\begin{equation} 
m_{\nu_{ij}} = \frac{1}{\left|{\cal M}_{\rm NMSSM}\right|} 
\left[ 
\mu_i \mu_j {\cal F}  
+ ( \mu_i \Lambda_j + \mu_j \Lambda_i ) {\cal G}  
+ \Lambda_i \Lambda_j {\cal H} 
\right],   
\label{MatrixForm} 
\end{equation} 
where $\left|{\cal M}_{\rm NMSSM}\right|$ is the determinant of matrix
(\ref{NMSSMmass}), $\Lambda_i= \langle s \rangle (\lambda_i+\lambda
\frac{v_i}{v_d})$ and:
\begin{eqnarray} 
{\cal F} & = & \lambda^2 v_u^2 M_1 M_2 + {\cal X},  
\nonumber  \\    
{\cal G} & = & {\cal X} + (\lambda v_d - \cos^2 \beta \sum_{i=1}^3
\lambda_i v_i) \ {\cal Y},
\label{functions}  \\   
{\cal H} & = & {\cal X} + 2 \cos^2 \beta (\lambda v_d - \sum_{i=1}^3
\lambda_i v_i) \ {\cal Y},
\nonumber    
\end{eqnarray}
with,
\begin{eqnarray} 
{\cal X} & = & 2 \cos^2 \beta \kappa \langle s \rangle M_Z^2 
( \cos^2 \theta_W M_1 + \sin^2 \theta_W M_2 ),
\nonumber  \\ 
{\cal Y} & = & \frac{v_u}{\langle s \rangle} M_Z^2 
( \cos^2 \theta_W M_1 + \sin^2 \theta_W M_2 ).
\nonumber  
\end{eqnarray}
One should notice that the neutrino mass matrix (\ref{MatrixForm}) arises
entirely at the tree level.

The emergence of the specific mass matrix structure of
Eq.~(\ref{MatrixForm}) at tree level is the primary result of our
analysis. Accordingly to this particular structure, if the two
(effective) quantities $\mu_i$ and $\Lambda_i$ (characteristic of the
NMSSM) take simultaneously non-vanishing values, then the mass matrix
ceases to be of rank 1, even though the determinant is still zero.  In
this situation, we get two non-zero neutrino mass eigenvalues.  It is
worth comparing the situation with what happens in the MSSM with
bilinear \rpv violation. In the latter case, we get a similar kind of
analytic structure of the mass matrix from the simultaneous
consideration of the $\mu_i$ as well as the corresponding soft $B_i$
terms.  While the $\mu_i \mu_j$ contributions originate at tree level,
the $\mu_i B_j$ and $B_i B_j$ contributions arise at one-loop order
through Grossman-Haber diagrams \cite{GroHab} which proceed through
slepton-Higgs and neutrino-neutralino mixings (for a series of
analysis in a three-flavour framework, see \cite{asmaa}). The
Grossman-Haber loops would still contribute in our scenario, but now
that we have two tree level masses, those loop-suppressed
contributions are not so crucial for generating a viable neutrino mass
spectrum.
 
{\bf Neutrino mass eigenvalues at tree level:} 
The eigenvalues of the effective neutrino mass matrix (\ref{MatrixForm})
turn out to be 
$\{ 0 , m_{\nu}^- , m_{\nu}^+ \}$ with,
\begin{eqnarray}
& &m_{\nu}^\pm   =  \frac{1}{2\left|{\cal M}_{\rm NMSSM}\right|} 
\left[ 
\left(\sum_{i=1}^{3} \mu_i^2\right) {\cal F}     
+ \left(\sum_{i=1}^{3} \Lambda_i^2\right) {\cal H}     
+ 2 \left(\sum_{i=1}^{3} \mu_i \Lambda_i\right) {\cal G}     
\right. \nonumber \\
& &\left.  \pm    
\left\{ 
\left[  
\left(\sum_{i=1}^{3} \mu_i^2\right) \left(\sum_{i=1}^{3} \Lambda_i^2\right)  
- \left(\sum_{i=1}^{3} \mu_i \Lambda_i\right)^2  
\right] {\cal I} 
+ 
\left[ 
\left(\sum_{i=1}^{3} \mu_i^2\right) {\cal F}     
+ \left(\sum_{i=1}^{3} \Lambda_i^2\right) {\cal H}     
+ 2 \left(\sum_{i=1}^{3} \mu_i \Lambda_i\right) {\cal G}     
\right]^2     
\right\}^{\frac{1}{2}} 
\right] \nonumber, \\ 
\label{Eigen} 
\end{eqnarray} 
where ${\cal F},{\cal G},{\cal H}$ are defined in
Eq.~(\ref{functions}), and ${\cal I} = 4({\cal G}^2 - {\cal F} {\cal
H})$.

Note that for either all $\mu_i = 0$ or all $\Lambda_i = 0$, the
eigenvalue $m_{\nu}^-$ vanishes as expected since in this limit we
recover the rank 1 form. An inspection of the form of
Eq.~(\ref{Eigen}) reveals that the coefficient of ${\cal I}$ can be
written as $\sum_{i\neq j} (\mu_i \Lambda_j -
\mu_j \Lambda_i)^2$, which indicates that the misalignment between $\mu_i$
and $\Lambda_i$ is crucial in creating a non-vanishing $m_\nu^{-}$.  

Therefore, the condition for generating two non-vanishing and
non-degenerate eigenvalues is to ensure $\mu_i \neq 0$ and $\Lambda_i
\neq 0$ simultaneously. In other words, to achieve two non-zero eigenvalues, 
$\mu_i$ has to be non-zero always, but we can go to a basis of
$L_\alpha$ fields where $v_i = 0$ (then generally $\lambda_i \neq 0$),
or to the other extreme where $\lambda_i = 0$ (but $v_i \neq 0$), or
any other basis in between basically maintaining $\Lambda_i \neq
0$. On the contrary, if $\mu_i=0$, only one neutrino eigenstate gets a
mass different from zero, as was also found by the authors of
Ref.~\cite{rpNMSSM1}.  But all the scenarios we considered in this
paper, according to the above arguments, will yield two non-zero
neutrino masses at tree level. We mention that within the first
scenario, a rotation on the $L_\alpha$ fields has already been
performed, whereas in the third one, no rotation is possible.

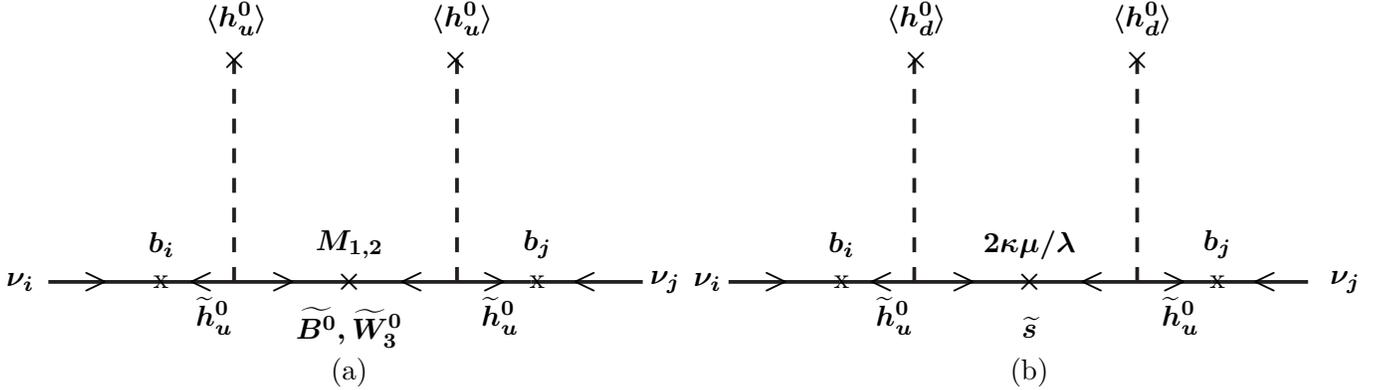
\begin{figure}[t]\unitlength1mm
\SetScale{2.8}
\begin{boldmath}
\begin{tabular}{c c c}
\begin{picture}(80,80)(0,-10)
\Line(0,0)(15,0)\Line(65,0)(15,0)
\Line(80,0)(65,0)
\DashLine(25,0)(25,30){2}
\Text(22,-4.5)[c]{$\widetilde  {h}^0_{u}$}
\DashLine(55,0)(55,30){2}
\Text(60,-4.5)[c]{$\widetilde  {h}^0_{u}$}
\Text(52.4,29.6)[l]{$\times$}
\Text(55,35)[c]{$\langle {h}^0_{u} \rangle $}
\Text(23,29.6)[l]{$\times$}
\Text(25,35)[c]{$\langle {h}^0_{u} \rangle $}
\Text(73,0)[r]{$<$}
\Text(61,0)[r]{$>$}
\Text(-2,0)[r]{$\nu_i$}
\Text(8,0)[r]{$>$}
\Text(22,0)[r]{$<$}
\Text(80,0)[l]{$\nu_j$}
\Text(40,-6)[c]{$\widetilde{B^0},\widetilde{W}^0_3$}
\Text(40,-12)[c]{(a)}
\Text(40,5)[c]{$M_{1,2}$}
\Text(33,0)[r]{$>$}
\Text(50,0)[r]{$<$}
\Text(40,0)[c]{$\times$}
\Text(15,0)[c]{x}
\Text(65,0)[c]{x}
\Text(15,5)[c]{$b_i$}
\Text(65,5)[c]{$b_j$}
\end{picture}
&\strut\hspace*{2mm}&
\begin{picture}(80,80)(0,-10)
\Line(0,0)(15,0)\Line(65,0)(15,0)
\Line(78,0)(65,0)
\DashLine(25,0)(25,30){2}
\Text(22,-4.5)[c]{$\widetilde  {h}^0_{u}$}
\DashLine(55,0)(55,30){2}
\Text(60,-4.5)[c]{$\widetilde  {h}^0_{u}$}
\Text(52.6,29.6)[l]{$\times$}
\Text(55,35)[c]{$\langle {h}^0_{d} \rangle $}
\Text(23.2,29.6)[l]{$\times$}
\Text(25,35)[c]{$\langle {h}^0_{d} \rangle $}
\Text(73,0)[r]{$<$}
\Text(61,0)[r]{$>$}
\Text(-1,0)[r]{$\nu_i$}
\Text(8,0)[r]{$>$}
\Text(22,0)[r]{$<$}
\Text(80,0)[l]{$\nu_j$}
\Text(40,-6)[c]{$\widetilde{s}$}
\Text(33,0)[r]{$>$}
\Text(50,0)[r]{$<$}
\Text(40,5)[c]{$2\kappa \mu/\lambda$}
\Text(40,0)[c]{$\times$}
\Text(15,0)[c]{x}
\Text(65,0)[c]{x}
\Text(15,5)[c]{$b_i$}
\Text(65,5)[c]{$b_j$}
\Text(40,-12)[c]{(b)}
\end{picture}
\end{tabular}
\end{boldmath}
\caption{\small \sf Tree level Feynman diagrams in the \rpv NMSSM generating
Majorana neutrino masses proportional to $b_i b_j$.  The effective
\rpv bilinear parameter $b_i$ stands for either $\mu_i$ or $\lambda_i
\langle s \rangle$.  The mass-dimensional couplings appearing at the
two vertices are of the type $m=M_Z t(\theta_W) \sin \beta$
[$t(\theta_W)=\sin \theta_W$ for $\tilde{B}^0$ and $t(\theta_W)= -
\cos \theta_W$ for $\tilde{W}^0_3$] in (a) and $m=-\lambda v_d$ in
(b).  A cross indicates either a mass insertion or a vev. The arrows
show the flow of momentum for the associated propagators.}
\protect\label{fig:FIRST}
\end{figure}

\begin{figure}[t]\unitlength1mm
\SetScale{2.8}
\begin{boldmath}
\begin{tabular}{c c c}
\begin{picture}(80,80)(0,-10)
\Line(0,0)(15,0)\Line(65,0)(15,0)
\Line(78,0)(65,0)
\DashLine(25,0)(25,30){2}
\Text(18,5)[c]{$\lambda_i v_{u}$}
\DashLine(55,0)(55,30){2}
\Text(61,5)[c]{$\lambda_j v_{u}$}
\Text(52.6,29.6)[l]{$\times$}
\Text(55,35)[c]{$\langle {h}^0_{u} \rangle $}
\Text(22.9,29.6)[l]{$\times$}
\Text(25,35)[c]{$\langle {h}^0_{u} \rangle $}
\Text(73,0)[r]{$<$}
\Text(-1,0)[r]{$\nu_i$}
\Text(8,0)[r]{$>$}
\Text(80,0)[l]{$\nu_j$}
\Text(40,-6)[c]{$\widetilde{s}$}
\Text(33,0)[r]{$<$}
\Text(50,0)[r]{$>$}
\Text(40,5)[c]{$2\kappa \mu/\lambda$}
\Text(40,0)[c]{$\times$}
\Text(40,-12)[c]{(a)}
\end{picture}
&\strut\hspace*{2mm}&
\begin{picture}(80,80)(0,-10)
\Line(0,0)(15,0)\Line(65,0)(15,0)
\Line(78,0)(65,0)
\DashLine(25,0)(25,30){2}
\DashLine(55,0)(55,30){2}
\Text(60,-4.5)[c]{$\widetilde  {h}^0_{u}$}
\Text(52.6,29.6)[l]{$\times$}
\Text(55,35)[c]{$\langle {h}^0_{d} \rangle $}
\Text(23.2,29.6)[l]{$\times$}
\Text(25,35)[c]{$\langle {h}^0_u \rangle $}
\Text(73,0)[r]{$<$}
\Text(61,0)[r]{$>$}
\Text(-1,0)[r]{$\nu_i$}
\Text(8,0)[r]{$>$}
\Text(80,0)[l]{$\nu_j$}
\Text(40,-6)[c]{$\widetilde{s}$}
\Text(40,0)[r]{$<$}
\Text(65,0)[c]{x}
\Text(18,5)[c]{$\lambda_i v_u$}
\Text(65,5)[c]{$b_j$}
\Text(40,-12)[c]{(b)}
\end{picture}
\end{tabular}
\end{boldmath}
\caption{\small \sf Tree level Feynman diagrams in the \rpv NMSSM generating
Majorana neutrino masses proportional to $\lambda_i \lambda_j v_u^2$ and
$\lambda_i v_u b_j$. The effective \rpv bilinear parameter $b_i$, as in Figure
1, stands for either $\mu_i$ or $\lambda_i \langle s \rangle$. The
mass-dimensional coupling appearing at the right vertex in (b) is of the type
$m= - \lambda v_{d}$.}  \protect\label{fig:SECOND}
\end{figure}
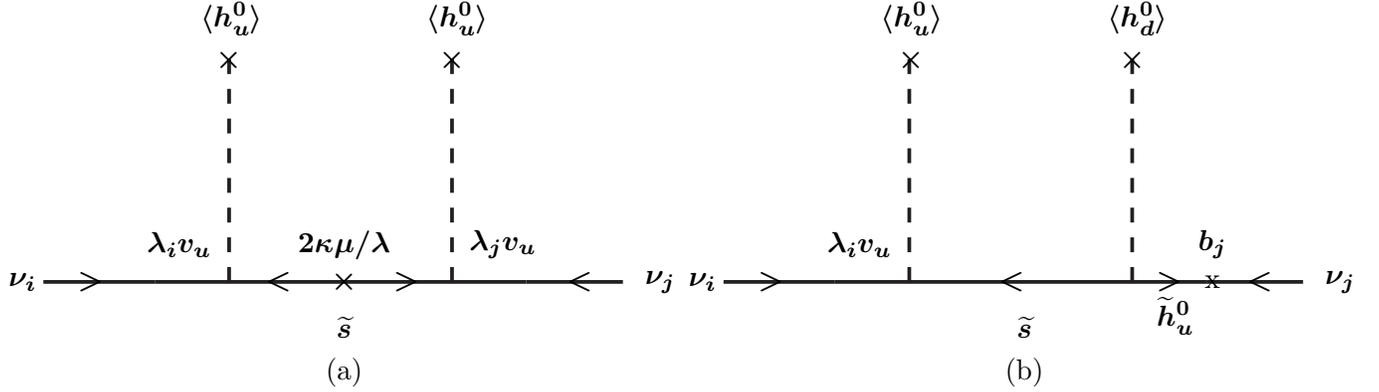

In Figures (\ref{fig:FIRST}) and (\ref{fig:SECOND}), we present the
Feynman diagrams contributing to the Majorana neutrino mass
(\ref{MatrixForm}). All these diagrams proceed through the tree level
exchange of the neutralinos (gauginos, higgsinos and singlino).
In these figures, we have considered the basis corresponding to
$v_i=0$ for simplicity.

\section{Numerical results} 
\label{Numeric}
Thus the present model predicts a hierarchical neutrino mass spectrum
at tree level. This hierarchical pattern could be modified by the loop
level contributions to neutrino masses. At least, the massless state
acquires a mass once the loop contributions are considered.

It is not our aim to give detailed numerical fits in this short paper. We
would just like to numerically demonstrate that our scenario can reproduce the
neutrino data from the tree level neutrino mass matrix, with a choice of NMSSM
parameters that corroborate the $\mu$-naturalness.

As an example, we take the following NMSSM parameters:
\begin{equation} 
\lambda=0.4, \ \kappa=0.3, \ \mu=-200 ~\mbox{GeV}, \ \tan \beta=30,  
\ M_1=350 ~\mbox{GeV}, \ M_2=500 ~\mbox{GeV} . 
\label{paramA} 
\end{equation}  
These parameters satisfy the general NMSSM constraints described
below.  These constraints are not expected to be significantly
modified by the presence of \rpv interactions in superpotential
(\ref{Wgeneric}) as we work under the assumption $|\mu_i/\mu| \ll 1$
and $|\lambda_i/\lambda| \ll 1$.

The $\mu$-naturalness forces one to restrict to $\langle s \rangle
\lesssim 10 ~\mbox{TeV}$, which translates into the condition $|\mu|
[\mbox{GeV}] \times 10^{-4} \lesssim \lambda$.  Furthermore, the
absence of Landau singularities, for $\lambda$, $\kappa$, the top and
bottom Yukawa coupling constants below the GUT energy scale, imposes
\cite{NMpheno} the typical bounds: $\lambda
\lesssim 0.75$, $|\kappa| \lesssim 0.65$ and $1.7 \lesssim \tan \beta \lesssim
54$. Finally, the LEP bound on the lightest chargino mass, namely
$m_{\tilde \chi_1^+}>103.5~\mbox{GeV}$ \cite{LEPchargino}, translates
into $|\mu| \gtrsim 100 ~\mbox{GeV}$.

Together with the values in Eq.~(\ref{paramA}), we take the following \rpv
effective couplings,
\begin{eqnarray}
\mu_1=1 \cdot 10^{-4} ~\mbox{GeV}, \ \mu_2=2 \cdot 10^{-4} ~\mbox{GeV}, 
\ \mu_3=2 \cdot 10^{-4} ~\mbox{GeV} \nonumber \\  
\Lambda_1 / \langle s \rangle =1.9 \cdot 10^{-5}, \ 
\Lambda_2 / \langle s \rangle =1.4 \cdot 10^{-5}, \ 
\Lambda_3 / \langle s \rangle =1.5 \cdot 10^{-5}.   
\label{paramB} 
\end{eqnarray}  
This set of parameters yield the following three neutrino mass eigenvalues at 
tree level: 
\begin{equation} 
m_{\nu_1} = 0, \ m_{\nu_2} = 0.0095 ~\mbox{eV}, 
\ m_{\nu_3} = 0.058 ~\mbox{eV}. 
\label{Realistic} 
\end{equation}
These values are in agreement with the three-flavour analyzes including
results from solar, atmospheric, reactor and accelerator oscillation
experiments which lead to ($4 \sigma$ level): $6.8 \leq \Delta m_{21}^2 \leq
9.3 \ [10^{-5} \mbox{eV}^2]$ and $1.1 \leq \Delta m_{31}^2 \leq 3.7 \ [10^{-3}
\mbox{eV}^2]$ \cite{Valle}.  Besides, the neutrino mass eigenvalues in
(\ref{Realistic}) satisfy the bound extracted from WMAP and 2dFGRS galaxy
survey (depending on cosmological priors): $\sum_{i=1}^3 m_{\nu_i} \lesssim
~0.7~\mbox{eV}$ \cite{cosmobound}. Finally, these eigenvalues are perfectly
compatible with the limits extracted from the tritium beta decay experiments
($95 \% \ {\rm C.L.}$): $m_\beta \leq 2.2 \ \mbox{eV} \ \mbox{[Mainz]}$ and
$m_\beta \leq 2.5 \ \mbox{eV} \ \mbox{[Troitsk]}$ \cite{absmass}, this
effective mass being defined as $m^2_\beta = \sum_{i=1}^3 |U_{ei}|^2
m_{\nu_i}^2$ where $U_{ei}$ is the leptonic mixing matrix.

Although we have chosen a particular set of input parameters for
illustration, solutions exist over a wide range of parameter space.
More realistic estimates can be obtained by switching on the soft
scalar terms $B_i \tilde
\ell_i h_u$ + h.c.  plus the superpotential terms $\lambda_{ijk} L_i L_j
E_k^c$ and $\lambda'_{ijk} L_i Q_j D_k^c$. All these terms contribute to the
neutrino mass matrix at one-loop order. A combined fit of all these parameters
is beyond the scope of this paper.

\section{Conclusion} 
\label{Conclusion}
In the NMSSM, which is a gauge singlet extension of the MSSM
addressing the $\mu$-naturalness, {\em two non-vanishing neutrino mass
eigenvalues can arise at tree level} when the lepton number violating
bilinear terms $\mu_i L_i H_u$ are present. One can then explain the
neutrino oscillation data without essentially depending on the
loop-generated masses which otherwise bring in more uncertainties from
unknown dynamics.  This result is in contrast with any other
supersymmetric \rpv scenario, as those scenarios do not generate more
than one massive neutrino eigenstate at tree level (except the
scenario proposed in \cite{Munoz} where 3 right-handed neutrinos are
added to the field content).

\vskip 5pt
{\bf Acknowledgments:} G.~M. thanks U.~Ellwanger for fruitful
discussions.  A.~A. acknowledges the {\it Agence Nationale Pour la
Recherche} (ANR), project NEUPAC. G.~B. acknowledges hospitality at
CFTP, Instituto Superior T\'ecnico, Lisbon, when this work was
initiated, and ICTP, Trieste, during the final phase of the
work. G.~M. acknowledges support from a \textit{Marie Curie}
Intra-European Fellowships (under contract MEIF-CT-2004-514138),
within the 6th European Community Framework Program.

\end{document}